\documentclass{PoS}

\newcommand\as{\alpha_{\mathrm{S}}} 
\newcommand\f[2]{\frac{#1}{#2}} 
\def\beq{\begin{equation}} 
\def\eeq{\end{equation}} 
\def\beeq{\begin{eqnarray}} 
\def\eeeq{\end{eqnarray}} 
 
\def\to{\rightarrow}
 
\def\nn{\nonumber}

\def\ms{${\overline {\rm MS}}$}
\def\bqt{{\bf q_T}}

\def\ep{\epsilon}

\def\qi{\{q_i\}}
\title{Transverse-momentum resummation and the structure of hard factors at the NNLO}

\ShortTitle{Transverse-momentum resummation and the structure of hard factors at the NNLO}

\author{\speaker{Leandro Cieri}\\
        Universita di Roma " La Sapienza"\\
        E-mail: \email{Cieri@roma1.infn.it}}

\abstract{
In this proceeding we consider QCD radiative corrections to the production of
colourless high-mass systems in hadron collisions.
At small transverse momentum the logarithmically-enhanced
contributions can be organized to all 
perturbative orders by a universal resummation formula that depends on a single
process-dependent hard factor. We show that the hard factor is directly related to the
all-order virtual amplitude of the corresponding partonic process by a universal
(process independent) formula, which we explicitly evaluate up to two-loop level.
Once the next-to-next-to-leading order (NNLO) scattering amplitude is available,
the corresponding hard factor
is directly determined. It can be used in fully-exclusive perturbative
calculations (\textit{via} $q_T$ subtraction formalism) up to NNLO, in resummed
calculations at full next-to-next-to-leading logarithmic (NNLL) accuracy, and
also, it's a necessary ingredient to the next subsequent logarithmic order (N$^3$LL).
}

\FullConference{ Loops and Legs in Quantum Field Theory - LL 2014,\\
                 27 April - 2 May 2014 \\
                 Weimar, Germany }

\begin{document}

\section{Introduction}
We consider the inclusive hard-scattering reaction
\begin{equation}
h_1(p_1)+h_2(p_2)\to F(\{q_i\})+X\, ,
\label{class}
\end{equation}
where the collision of the two hadrons $h_1$ and $h_2$ with momenta $p_1$ and
$p_2$ produces the observed final state $F$, accompanied by an arbitrary and undetected final state $X$.
The triggered final state $F$ is a generic system of one or more
{\em colourless} particles,
such as lepton pairs (produced by Drell--Yan (DY) mechanism), photon pairs, 
vector bosons, Higgs boson(s), and so forth. The momenta of these final state particles are denoted by $q_1$,$q_2$...$q_n$.
The system $F$ has {\em total} invariant mass $M^2=(q_1+q_2+...q_n)^2$, transverse momentum $\bqt$ and rapidity $y$. We employ $\sqrt{s}$ to denote the centre-of-mass energy 
of the colliding hadrons, which are treated in the massless
approximation ($s=(p_1+p_2)^2=2p_1\cdot p_2$).

It is possibile to calculate the transverse-momentum $(q_T)$ cross section for the process in Eq.~(\ref{class}) by using perturbative QCD. In the small-$q_T$ region (roughly, in the region where
$q_T \ll M$) the convergence of the fixed-order perturbative expansion
in powers of the QCD coupling $\as$ is spoiled by 
the presence of large logarithmic 
terms of the type $\ln^n(M^2/q_T^2)$. We can recover the predictivity of perturbative QCD performing the summation of these logarithmically-enhanced contributions to all order in $\as$ 
\cite{Dokshitzer:hw,Collins:1984kg,Kodaira:1981nh}. 

If the final state $F$ is colourless, the large
logarithmic contributions to the $q_T$ cross section can be systematically
resummed to all perturbative orders, and 
the structure of the resummed calculation can be arranged in a
{\em process-independent} form 
\cite{Dokshitzer:hw,Kodaira:1981nh, Catani:2000vq, Catani:2010pd}.
Starting from the resummation formula for the DY process \cite{Collins:1984kg},
two additional steps were needed to arrive at the process-independent version of the formalism: the understanding of the all-order process-independent structure of the Sudakov form factor
(through the factorization of a single process-dependent hard factor)
\cite{Catani:2000vq}, and the complete generalization to processes that are initiated by
the gluon fusion mechanism \cite{Catani:2010pd}.

The all-order process-independent form of the resummed calculation
has a factorized structure, whose resummation factors are 
(see Sect.~\ref{sec:resu}) the (quark and gluon) Sudakov form factor,
process-independent {\em collinear} factors and a process-dependent
{\em hard} or, more precisely (see Sect.~\ref{sec:hardvirtual}), hard-virtual  
factor. These factors (which are a set of perturbative functions
whose perturbative {\em resummation coefficients} are computable order-by-order
in $\as$) control the resummation of the logarithmic contributions.
 The perturbative coefficients of the Sudakov form factor
are known, since some time
\cite{Kodaira:1981nh, Davies:1984hs,deFlorian:2000pr,deFlorian:2001zd},
up to the second order in $\as$, and the third-order coefficient 
$A^{(3)}$ (which is necessary to explicitly perform resummation up to the
next-to-next-to-leading logarithmic (NNLL) accuracy) is also known 
\cite{Becher:2010tm}.
The next-to-next-to-leading order (NNLO) QCD calculation of the $q_T$ cross
section (in the small-$q_T$ region) has been done in analytic form for two 
benchmark processes, namely, SM Higgs boson production \cite{Catani:2011kr}
and the DY process \cite{Catani:2012qa}. The results of 
Refs.~\cite{Catani:2011kr, Catani:2012qa} provide us with the
complete 
knowledge
of the process-independent {\em collinear} resummation coefficients up to the
second order in $\as$, and with the explicit expression of the hard coefficients
for these two specific processes. As shown in Ref.~\cite{Catani:2013tia}, the hard factor (which is process dependent) has an universal (process-independent) structure. The universality
structure of the factorization formula has a {\em soft} (and collinear) origin, and
it is closely
(though indirectly) related to the universal structure of the infrared divergences
\cite{Catani:1998bh}
of the scattering amplitude.
This process-independent structure of the hard-virtual term, which generalizes the 
next-to-leading order (NLO) results of Ref.~\cite{deFlorian:2001zd}, is valid to
all perturbative orders \cite{Catani:2013tia}. 
The NNLO universal formula for the hard-virtual term completes the $q_T$ resummation
formalism in explicit form up to full NNLL+NNLO accuracy.
This permits direct applications to NNLL+NNLO resummed calculations for any 
processes
of the class in Eq.~(\ref{class}) (provided the corresponding NNLO amplitude is
known), as already done for the cases of SM Higgs boson 
\cite{Bozzi:2005wk}
and DY \cite{Bozzi:2010xn,Guzzi:2013aja} production. The NNLO information of the $q_T$ resummation formalism is also relevant in the
context of {\em fixed order} calculations. Indeed, it enables to carry out
fully-exclusive NNLO calculations by applying the $q_T$ {\em subtraction
formalism} of Ref.~\cite{Catani:2007vq}
(the subtraction counterterms of the formalism follow \cite{Catani:2007vq}
from the fixed-order expansion of the $q_T$ resummation formula, 
as in Sect.~2.4 of Ref.~\cite{Bozzi:2005wk}). 
The $q_T$ subtraction
formalism has been applied 
to the NNLO computation of Higgs boson \cite{Catani:2007vq,Grazzini:2008tf}
and vector boson production \cite{Catani:2009sm},
associated production of the Higgs boson with a $W$ boson \cite{Ferrera:2011bk}, 
diphoton production \cite{Catani:2011qz}, $Z\gamma$ production \cite{Grazzini:2013bna} and $ZZ$ production \cite{Cascioli:2014yka}.
The computations of 
Refs.~\cite{Catani:2007vq,Grazzini:2008tf,Catani:2009sm,Ferrera:2011bk}
were based on the specific calculation  of the NNLO hard-virtual coefficients
of the corresponding processes \cite{Catani:2011kr,Catani:2012qa}.
The computations of Refs.~\cite{Catani:2011qz, Grazzini:2013bna, Cascioli:2014yka}
used the NNLO hard-virtual coefficients that are determined by applying
the universal form of the hard-virtual term that is derived in~\cite{Catani:2013tia} and illustrated in the present proceeding.

Transverse-momentum resummation can equivalently be reformulated by 
using $q_T$-dependent partonic distributions (see, e.g., 
Refs.~\cite{Becher:2010tm, qtdep}). 
The explicit NNLO results for the process-independent collinear coefficients 
\cite{Catani:2007vq, Catani:2009sm, Catani:2011kr, Catani:2012qa} 
and for the structure of the hard-virtual
coefficients \cite{Catani:2013tia} have been confirmed by the fully-independent
computation of Ref.~\cite{Gehrmann:2012ze}, 
which uses the formalism of Ref.~\cite{Becher:2010tm}.

\vspace*{-1mm}
\section{Small-$q_T$ resummation}
\label{sec:resu}
We consider the inclusive-production process in Eq.~(\ref{class}), and we introduce the 
corresponding {\em fully} differential cross section\footnote{In this section we briefly recall the
formalism of transverse-momentum resummation in impact parameter space 
\cite{Dokshitzer:hw,Kodaira:1981nh, deFlorian:2000pr, Catani:2000vq,
Catani:2010pd}. We closely follow the notation of Ref.~\cite{Catani:2010pd}
(more details about our notation can be found therein).}
\begin{equation}
\label{diffxs}
\f{d\sigma_F}{d^2{\bqt} \;dM^2 \;dy \;d{\Omega}} 
\,(p_1, p_2;\bqt,M,y,
\Omega )
\;\;,
\end{equation}
which depends on the total momentum of the system $F$
(i.e. on the variables $\bqt, M, y$). To evaluate the $\bqt$ dependence of the differential cross section
in Eq.~(\ref{diffxs}) within QCD perturbation theory, 
we first propose the following decomposition:
\begin{equation}
\label{Fdec}
d\sigma_F =
d\sigma_F^{({\rm sing})} +
\; d\sigma_F^{({\rm reg})}
\;\;.
\end{equation}
The two last terms in the right-hand side already include the  convolutions of
partonic cross sections and the scale-dependent parton distributions 
$f_{a/h}(x,\mu^2)$  ($a=q_f, {\bar q}_f, g$ is the parton label) of the
colliding hadrons. We use parton densities as defined in 
the \ms\ factorization scheme, and $\as(q^2)$ is
the QCD running coupling in the \ms\ renormalization scheme. 
The partonic cross sections that enter the singular component (the first term
in the right-hand side of Eq.~(\ref{Fdec})) contain all 
the contributions that are enhanced
(or `singular') at small $q_T$. These contributions are proportional to
$\delta^{(2)}(\bqt)$ or to large logarithms of the type 
$\f{1}{q_T^2}\ln^m (M^2/q_T^2)$. The partonic cross sections of the second term in the right-hand side of Eq.~(\ref{Fdec}) are regular (i.e. free of logarithmic terms)
order-by-order in perturbation theory as 
$q_T \to 0$.
In the following we focus on
the singular component, $d\sigma_F^{({\rm sing})}$,
which has an universal all-order
structure. The corresponding resummation formula is written as \cite{Dokshitzer:hw,Catani:2000vq,Catani:2010pd}
\begin{eqnarray}
\label{qtycross}
\f{d\sigma_F^{({\rm sing})}(p_1, p_2;\bqt,M,y,{\Omega} )}{d^2{\bqt} \;dM^2 \;dy \;d{\Omega}} 
&=&\f{M^2}{s}\sum_{c=q,{\bar q},g} \;
\left[d\sigma_{c{\bar c},F}^{(0)}\right]
\int \f{d^2{\bf b}}{(2\pi)^2} \;\, e^{i {\bf b}\cdot \bqt} \;
  S_c(M,b)\nn \\
 \;\;\;\; &\times& \;
\sum_{a_1,a_2} \,
\int_{x_1}^1 \f{dz_1}{z_1} \,\int_{x_2}^1 \f{dz_2}{z_2} 
\; \left[ H^F C_1 C_2 \right]_{c{\bar c};a_1a_2}
\;f_{a_1/h_1}
\;f_{a_2/h_2} \;
\;, 
\end{eqnarray}
where $b_0=2e^{-\gamma_E}$
($\gamma_E=0.5772\dots$ is the Euler number) is a numerical coefficient,
and the kinematical variables $x_1= \f{M}{\sqrt s} \;e^{+y}$ and $x_2=\f{M}{\sqrt s} \;e^{-y}$.
The function $S_c(M,b)$ is the Sudakov form factor, which is universal (process
independent)~\cite{Catani:2000vq}: it only depends on the type ($c=q$ or $c=g$) of colliding partons, and it resums the logarithmically-enhanced contributions of the form $\ln M^2b^2$ (the region $q_T \ll M$ corresponds to $Mb \gg 1$ in impact parameter space). The all-order expression of $S_c(M,b)$ is \cite{Collins:1984kg}
\begin{equation}
\label{formfact}
S_c(M,b) = \exp \left\{ - \int_{b_0^2/b^2}^{M^2} \frac{dq^2}{q^2} 
\left[ A_c(\as(q^2)) \;\ln \frac{M^2}{q^2} + B_c(\as(q^2)) \right] \right\} 
\;\;,
\end{equation}
where $A_c(\as)$ and $B_c(\as)$ are perturbative series in $\as$.
The perturbative coefficients $A^{(1)}_c, B^{(1)}_c, A^{(2)}_c$
\cite{Kodaira:1981nh},  $B^{(2)}_c$ 
\cite{Davies:1984hs, deFlorian:2000pr, deFlorian:2001zd}
and $A^{(3)}_c$ \cite{Becher:2010tm} are explicitly known.

The Born level factor\footnote{The cross section at its corresponding {\em lowest
order} in $\as$.} $\left[ d\sigma_{c{\bar c}, \,F}^{(0)} \right]$ in Eq.~(\ref{qtycross})
is obviously process dependent, although its process dependence is elementary
(it is simply due to the Born level scattering amplitude of the partonic process 
$c{\bar c}\to F$). The remaining process dependence of Eq.~(\ref{qtycross})
is embodied in the `hard-collinear' factor $\left[ H^F C_1 C_2 \right]$.
This factor includes a process-independent part and a process-dependent part.
The structure of the process-dependent part is the main subject of the 
present proceeding.


In the case of processes that are initiated at the Born level
by the $q{\bar q}$ annihilation
channel ($c=q$), the symbolic factor $\left[ H^F C_1 C_2 \right]$ in 
Eq.~(\ref{qtycross}) has the following explicit form \cite{Catani:2000vq}
\begin{equation}
\label{what}
\left[ H^F C_1 C_2 \right]_{q{\bar q};a_1a_2}
 = H_q^F(x_1p_1, x_2p_2; {\Omega}; \as(M^2))
\;\, C_{q \,a_1}(z_1;\as(b_0^2/b^2)) 
\;\, C_{{\bar q} \,a_2}(z_2;\as(b_0^2/b^2)) \;\;,
\end{equation}
and the functions $H_q^F$ and $C_{q \,a}= C_{{\bar q} \,{\bar a}}$ have 
the perturbative expansion
\begin{eqnarray}
\label{hexp}
H_q^F(x_1p_1, x_2p_2; {\Omega}; \as) &=& 1+ \sum_{n=1}^\infty 
\left( \frac{\as}{\pi} \right)^n 
H_q^{F \,(n)}(x_1p_1, x_2p_2; {\Omega})
\;\;, \\
\label{cqexp}
C_{q \,a}(z;\as) &=& \delta_{q \,a} \;\,\delta(1-z) + 
\sum_{n=1}^\infty \left( \frac{\as}{\pi} \right)^n C_{q\, a}^{(n)}(z) \;\;.
\end{eqnarray}
The function $H_q^F$ is process dependent, whereas the functions
$C_{q \,a}$ are universal (they only depend on the parton indices). The factorized structure in the right-hand side of Eq.~(\ref{what}) is based on the following fact: the scale of $\as$ is $M^2$ in the case of $H_q^F$, whereas
the scale is $b_0^2/b^2$ in the case of $C_{q \,a}$. The appearance of these two
different scales is essential \cite{Catani:2000vq} to disentangle the process
dependence of $H_q^F$ from the process-independent Sudakov form factor
($S_q$) and collinear functions ($C_{q \,a}$). In the case of processes that start at Born level by the 
gluon fusion channel ($c=g$), the physics of the small-$q_T$ cross section
has a richer structure, which is the consequence of collinear correlations
\cite{Catani:2010pd}
that are produced by the evolution of the colliding hadrons into gluon partonic
states
(the interested reader is referred to \cite{Catani:2010pd,Catani:2013tia}).
Despite its richer structure, it is possible to disentangle~\cite{Catani:2010pd} the process
dependence of $H_g^F$ from the process-independent Sudakov form factor
($S_c$) and collinear tensor functions ($C^{\mu\nu}_{g \,a}$) 
analogously to the case of the $q{\bar q}$ channel.

As a consequence of the renormalization-group symmetry (Eqs.(22)--(25), in Ref.~\cite{Catani:2013tia}),
the resummation factors $H^F$, $S_c$ and $C_{qa}$ are
not {\em separately} defined (and, thus, computable) in an unambiguous way.
Equivalently, each of these separate factors can be precisely defined only by
specifying
a {\em resummation scheme} \cite{Catani:2000vq}.
We choose the {\em hard scheme}, that is defined as follows.
The flavour off-diagonal coefficients $C^{(n)}_{ab}(z)$,
with $a\neq b$,
are `regular' functions of $z$ as $z\to 1$.
The $z$ dependence of 
the flavour diagonal coefficients $C^{(n)}_{qq}(z)$ and $C^{(n)}_{gg}(z)$
in Eqs.~(\ref{cqexp}) is instead due to both
`regular' functions and
`singular' distributions in the limit $z\to 1$. The 'singular' distributions are
$\delta(1-z)$ and the customary plus-distributions of the form 
$[(\ln^k(1-z))/(1-z)]_+\,$ ($k=0,1,2\dots$).
The {\em hard scheme} is the scheme in which,
order-by-order in perturbation theory, the coefficients $C^{(n)}_{ab}(z)$ with 
$n\geq 1$
do not contain any $\delta(1-z)$ term.
We highlight (see also Sect.~\ref{sec:hardvirtual}) that this definition directly 
implies that all the process-dependent virtual corrections to the Born level
subprocesses are embodied in the resummation coefficient $H_c^F$.

We note that the specification of the hard scheme (or any other scheme)
has sole practical purposes of presentation
(theoretical results can be equivalently presented, as actually done in 
Refs.~\cite{Catani:2011kr} and \cite{Catani:2012qa}, by explicitly parametrizing
the resummation-scheme dependence of the resummation factors).
The $q_T$ cross section, its all-order
resummation formula (\ref{qtycross}) and any consistent perturbative truncation
(either order-by-order in $\as$ or in classes of logarithmic terms) of the latter
\cite{Catani:2000vq, Bozzi:2005wk} are completely independent of the resummation
scheme.

The first-order coefficients $C^{(1)}_{ab}(z)$ are explicitly known
\cite{deFlorian:2000pr,Davies:1984hs,deFlorian:2001zd,Kauffman:1991cx}. 
The second-order process-independent collinear coefficients $C_{ab}^{(2)}(z)$ 
have been independently computed in 
Refs.~\cite{Catani:2007vq, Catani:2009sm, Catani:2011kr,Catani:2012qa}
and in Ref.~\cite{Gehrmann:2012ze} by using two completely different methods,
and the results of the two computations are in agreement.

The universality structure
of the process-dependent coefficients $H_c^F$ at NNLO and higher orders
(see Sect.~\ref{sec:hardvirtual})
is one of the main results that we are discussing in the present proceeding.

\vspace*{-2mm}
\section{Hard-virtual coefficients}
\label{sec:hardvirtual}

In the hard scheme that we are using, the hard-virtual coefficient contains all the information on the
process-dependent virtual corrections, 
and, therefore,  
we can show that 
$H^F$ can be related in a process-independent (universal) way to the multiloop virtual amplitude
${\cal M}_{c{\bar c}\to F}$ of the partonic process $c{\bar c}\to F$.

We consider the partonic {\em elastic}-production process
\begin{equation}
c({\hat p}_1)+ {\bar c}( {\hat p}_2)\to F(\{q_i\})\, ,
\label{partpro}
\end{equation}
where the two colliding partons with momenta ${\hat p}_1$ and ${\hat p}_2$
are either $c{\bar c}=gg$ or $c{\bar c}=q{\bar q}$ 
and $F(\{q_i\})$ is the triggered 
final-state system in Eq.~(\ref{class}). The loop scattering amplitude of the process in Eq.~(\ref{partpro}) contains ultraviolet (UV) and infrared (IR) singularities,
which are regularized in $d=4-2\epsilon$ space-time dimensions by using the customary scheme of
conventional dimensional regularization.
The renormalized all-loop amplitude of the generic 
process in Eq.~(\ref{partpro}) is denoted by ${\cal M}_{c{\bar c}\to F}$ and it 
has the perturbative (loop) expansion
\begin{eqnarray}
\lefteqn{
{\cal M}_{c{\bar c}\to F}({\hat p}_1, {\hat p}_2;\qi) \!
= \left( \as(\mu_R^2) \,\mu_R^{2\ep} \right)^k
\bigg\{
{\cal M}_{c{\bar c}\to F}^{\,(0)}({\hat p}_1, {\hat p}_2; \qi)}\nn\\
&\!&
+\left( \frac{\as(\mu_R^2)}{2\pi}\right) 
{\cal M}_{c{\bar c}\to F}^{\,(1)}({\hat p}_1, {\hat p}_2; \qi; \mu_R)
\!+ \sum_{n=3}^{\infty} \left(\frac{\as(\mu_R^2)}{2\pi}\right)^{\!\!n}
\!\!{\cal M}_{c{\bar c}\to F}^{\,(n)}({\hat p}_1, {\hat p}_2; \qi; \mu_R)
\bigg\} ,
\label{ampli}
\end{eqnarray}
where the value $k$ of the overall power of $\as$ depends on the specific
process.
The perturbative terms ${\cal M}_{c{\bar c}\to F}^{\,(l)}$
$(l=1,2,\dots)$
are UV finite, but they still depend on $\ep$ 
(although this dependence is not explicitly denoted in Eq.~(\ref{ampli}))
and, 
in particular, they are IR divergent as $\ep\to 0$.
The IR divergent contributions to the scattering amplitude have a universal
structure \cite{Catani:1998bh}, which is explicitly known at the 
one-loop \cite{ir1loop, Catani:1998bh},
two-loop \cite{Catani:1998bh, ggFF2} 
and three-loop \cite{FF3, 3loopsing} level for the class of processes in
Eq.~(\ref{partpro}).

In Ref.~\cite{deFlorian:2001zd} we can find the universal (process-independent) relation between the NLO hard-virtual coefficient $H^{F\,(1)}$ and the leading-order (LO) amplitude ${\cal M}_{c{\bar c}\to F}^{\,(0)}$ and to the IR finite part of the NLO amplitude ${\cal M}_{c{\bar c}\to F}^{\,(1)}$. The relation between $H^F_c$ and ${\cal M}_{c{\bar c}\to F}$ can be extended to NNLO and to higher-order levels \cite{Catani:2013tia}. This extension can be formulated and expressed in simple and general terms by introducing an 
auxiliary (hard-virtual) amplitude $\widetilde{\cal M}_{c{\bar c}\to F}$
that is directly obtained from ${\cal M}_{c{\bar c}\to F}$ in a universal
(process-independent) way\footnote{The interested reader is referred to \cite{Catani:2013tia}, where there are all the formulae to obtain $\widetilde{\cal M}_{c{\bar c}\to F}$.}. In practice, $\widetilde{\cal M}_{c{\bar c}\to F}$
is obtained from ${\cal M}_{c{\bar c}\to F}$ by removing its IR divergences and a
{\em definite} amount of IR finite terms. The (IR divergent and finite) terms
that are removed from ${\cal M}_{c{\bar c}\to F}$ originate from real emission
contributions to the cross section and, therefore, these terms and 
$\widetilde{\cal M}_{c{\bar c}\to F}$ {\em specifically}
depend on the
transverse-momentum cross section of Eq.~(\ref{diffxs}). The relation between $H_c^F$ and ${\cal M}_{c{\bar c}\to F}$ is based on an universal all-order factorization formula~\cite{Catani:2013tia} that emerges from
the factorization properties of soft (and collinear) parton radiation.
We have explicitly determined this relation up to the NNLO~\cite{Catani:2013tia}.
More precisely, we have shown~\cite{Catani:2013tia} that 
this relation is fully determined by the structure of IR singularities of the
all-order amplitude ${\cal M}_{c{\bar c}\to F}$ and by
renormalization-group invariance up to a {\em single} coefficient (of
{\em soft} origin) at each
perturbative order.

We can relate the subtracted amplitude $\widetilde{\cal M}_{c{\bar c}\to F}$ to the process-dependent resummation coefficients $H^F_c$ of Eqs.~(\ref{qtycross}) and (\ref{what}). For processes initiated by $q{\bar q}$ annihilation
(see Eqs.~(\ref{what}) and (\ref{hexp})), 
the {\em all-order} coefficient $H^F_q$ can be written as
\begin{equation}
\label{Hq}
\as^{2k}(M^2) \,H^F_q(x_1p_1, x_2p_2; {\Omega};\as(M^2))
=\f{|\widetilde{\cal M}_{q{\bar q}\to F}(x_1p_1, x_2p_2;\qi)|^2}{|{\cal M}_{q{\bar
q}\to F}^{(0)}(x_1p_1, x_2p_2;\qi)|^2}\, ,
\end{equation}
where $k$ is the value of the overall power of $\as$ in the expansion of 
${\cal M}_{c{\bar c}\to F}$ (see Eq.~(\ref{ampli})).

The expression (\ref{Hq}) allows us the explicit computation of the process-dependent resummation coefficients 
$H^F_c$ for an arbitrary process of the class in Eq.~(\ref{class}). The computation of $H^F_c$ up to the NNLO is straightforward, provided the scattering amplitude ${\cal M}_{c{\bar c}\to F}$
of the corresponding partonic subprocess is available (known) up to the
NNLO (two-loop) level. 

Some examples (DY and Higgs boson production) are explicitly reported in 
Appendix~A of Ref.~\cite{Catani:2013tia}.
In particular, in Appendix~A of Ref.~\cite{Catani:2013tia}, we used Eq.~(\ref{Hq}), and we presented the explicit expression of the NNLO hard-virtual
coefficient $H^{\gamma \gamma (2)}_q$ for the process of diphoton production
\cite{Catani:2011qz}. Recently, Eq.~(\ref{Hq}) was used to obtain the
hard-virtual factor in the case of Higgs production in bottom quark
annihilation~\cite{Harlander:2014hya}, in order to calculate the transverse momentum distribution at NNLO+NNLL.

The same procedure that was applied to derive the universal formula for the
hard-virtual coefficient $H^F_c$ can be used within the related formalism of threshold
resummation \cite{Sterman:1986aj} for the {\em total} cross section.
The process-independent formalism of threshold resummation also involves a
corresponding process-dependent hard factor which has a
universality structure \cite{Catani:2013tia} that is analogous to the case of transverse-momentum
resummation. 
Recently, we also extended the threshold resummation results of Ref.~\cite{Catani:2013tia} 
to the next subsequent order (N$^3$LL) \cite{Catani:2014uta}. 
The general (process-independent) N$^3$LL results of Ref.~\cite{Catani:2014uta}
are based on the universality structure of the hard-virtual factor, and they
exploit the recent computation of the N$^3$LO Higgs boson cross section 
\cite{Anastasiou:2014vaa} within
the soft-virtual approximation. For the specific case of DY production we
confirm  \cite{Catani:2014uta} the soft-virtual N$^3$LO results of 
Ref.~\cite{Ahmed:2014cla}.

The results enumerated in this proceeding, with the knowledge of the other 
process-independent resummation coefficients, complete the $q_T$ resummation
formalism in explicit form up to full NNLL and NNLO accuracy
for all the processes in the class of Eq.~(\ref{class}).
This allows applications to NNLL+NNLO resummed calculations 
for {\em any} processes whose NNLO scattering amplitudes are available. Moreover, we have all the ingredients to implement the $q_T$ subtraction formalism \cite{Catani:2007vq} straightforwardly, to perform fully-exclusive NNLO computations for each of these processes.

\end{document}